\documentclass[fleqn,10pt]{wlscirep}
\usepackage[utf8]{inputenc}
\usepackage[T1]{fontenc}
\usepackage[round-mode=uncertainty,round-precision=1,separate-uncertainty=true]{siunitx}
\title{Analysis of complex excitation patterns using Feynman-like diagrams}

\author[1,3]{Louise Arno}
\author[1,2,3]{Desmond Kabus}
\author[1,3,*]{Hans Dierckx}
\affil[1]{Department of Mathematics, KU Leuven Campus Kortrijk (KULAK), Etienne Sabbelaan 53, 8500 Kortrijk, Belgium}
\affil[2]{Laboratory of Experimental Cardiology, Leiden University Medical Center (LUMC), Albinusdreef 2, 2333 ZA Leiden, The Netherlands}
\affil[3]{iSi Health, Institute of Physics-based Modeling for In Silico Health, KU Leuven, Oude Markt 13, 3000 Leuven, Belgium}
\affil[*]{dr.h.dierckx@gmail.com}

\begin{abstract}
Many extended chemical and biological systems self-organise into complex patterns that drive the medium behaviour in a non-linear fashion. An important class of such systems are excitable media, including neural and cardiac tissues. In extended excitable media, wave breaks can form rotating patterns and turbulence. However, the onset, sustaining and elimination of such complex patterns is currently incompletely understood. The classical theory of phase singularities in excitable media was recently challenged, as extended lines of conduction block were identified as phase discontinuities. Here, we provide a theoretical framework that captures the rich dynamics in excitable systems in terms of three quasiparticles: heads, tails, and pivots. We propose to call these quasiparticles `cardions'. In simulations and experiments, we show that these basic building blocks combine into at least four different bound states. By representing their interactions similarly to Feynman diagrams in physics, the creation and annihilation of vortex pairs are shown to be sequences of dynamical creation, annihilation, and recombination of the identified quasiparticles. We draw such diagrams for numerical simulations, as well as optical voltage mapping experiments performed on cultured human atrial myocytes (hiAMs).  Our results provide a new, unified language for a more detailed theory, analysis, and mechanistic insights of dynamical transitions in excitation patterns.
\end{abstract}
\begin{document}

\flushbottom
\maketitle
\thispagestyle{empty}

\section*{Introduction}
\label{sec:Intro}

Some of the most intriguing waves and patterns in nature arise from the spatial coupling between excitable elements.
Examples include the emergence of intelligence in neural tissue \cite{cannon_neurosystems_2014}, epidemic spread across a population \cite{smith_predicting_2002}, intracellular waves \cite{Lechleiter:1991}, chemical oxidation waves \cite{Rotermund:1990,Kapral:1995} and the emergence of cardiac arrhythmias in the heart muscle \cite{Allessie:1973,Gray:1998}.
Our work was motivated by the case of heart rhythm disorders, that had a global mortality rate of \qty{16}{\percent} in 2019, with increasing numbers every year. Hence, cardiac arrhythmias remain one of the largest causes of death worldwide \cite{WHO:2020}.

While the local excitation dynamics can be measured and modelled, a firm theoretical description at the mesoscale is still lacking. Hence, much effort is currently invested in numerical modelling \cite{Clayton:2011} and machine learning approaches \cite{trayanova_machine_2021}. Our aim is to complement such approaches with semi-analytical methods that give more insight in why patterns develop over time in a certain manner.

%Nonetheless, understanding these phenomena is challenge that is relevant both for science and the socio-economical impact.

\begin{figure*}[t]
    \centering
    \includegraphics{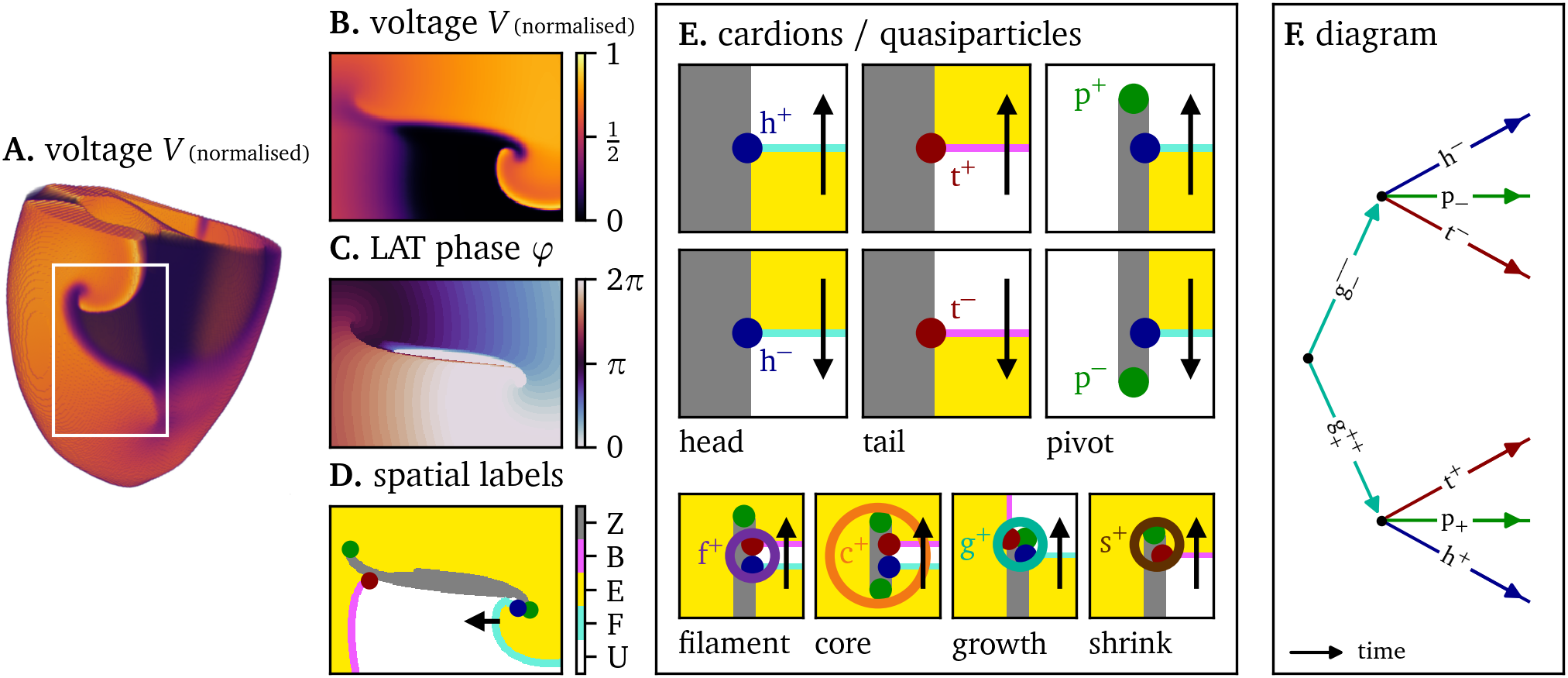}
    \caption[
        Conceptual framework for the analysis of excitation patterns using quasiparticles and Feynman-like diagrams.
    ]{
        Conceptual framework for the analysis of excitation patterns using quasiparticles and Feynman-like diagrams.
        \textbf{A.} A linear-core spiral wave\cite{BuenoOrovio:2008} in a numerical simulation on a human biventricular geometry, see the Methods section below. \textbf{B.} Numerical simulation\cite{BuenoOrovio:2008} in a two-dimensional domain, which could represent a pattern on the outer surface of the heart or in a cell culture.
        \textbf{C.} The linear phase based on local activation time (LAT) \eqref{phi_lin}.
        \textbf{D.} Spatial analysis of the observed pattern, showing excited ($\mathrm{E}$, yellow) and unexcited ($\mathrm{U}$, white) regions. They are separated by the wave fronts ($\mathrm{F}$, cyan), wave backs ($\mathrm{B}$, magenta), and phase defect lines ($\mathrm{Z}$, gray). The end points of these three different curves are topologically preserved as quasiparticles: heads (h, blue), tails (t, red), and pivots (p, green).
        \textbf{E.} Top: Depending on the orientation, we assign positive or negative charges $Q=\pm \frac{1}{2}$ to heads and tails, and charges $P = \pm \frac{1}{2}$ to pivots.
        Bottom: Several combinations of head, tail, and pivot are seen to travel together through the medium, akin to bound states in particle physics: the classical tip ($\mathrm{f} = \mathrm{h}+\mathrm{t}$), the spiral core ($\mathrm{c} = \mathrm{h}+\mathrm{t}+2\mathrm{p}$), and the newly identified phase defect growth sites ($\mathrm{g}=\mathrm{h}+\mathrm{t}+\mathrm{p}$), and shrinking sites ($\mathrm{s}=\mathrm{t}+\mathrm{p}$) seen during arrhythmogenesis. For an overview of all quantities, see Table \ref{tab:Particles}.
        \textbf{F.} We propose to keep track of the recombinations of building blocks over time using Feynman-like diagrams.
    }
    \label{fig:intro}
\end{figure*}

In this paper, we revise the classical mesoscale theory of excitation that has been used for cardiac arrhythmia analysis over the past decades \cite{Gray:1998, Clayton:2005, Clayton:2011, Christoph:2018}. Similar to historical developments in particle physics, we coin that the classical phase singularities in excitation are formed of more elementary quasiparticles. While this finer structure remains hidden in simple dynamical regimes, we show below in simulation and experiment that it becomes relevant during the processes of wave block, vortex initiation and recombination. Such insight is relevant for excitation patterns, since precisely those events need to be thoroughly understood if one  wants to predict or control the emergent dynamics in those systems.
A graphical outline of the analysis method presented in this paper can be found in Fig.~\ref{fig:intro}.

\begin{table}[ht]
\begin{center}
\caption[
    Overview of quasiparticles (cardions) in excitable media
]{
    Overview of quasiparticles (cardions) in excitable media, together with
    their charges, notation, composition, and colour code. In the extended
    notation, superscript denotes the head/tail charge $Q$ in units of
    $\frac{1}{2}$ and subscripts denote the pivot charge $P$. As only
    quasiparticles of the same charge bind, the simplified notation indicates
    this with a single superscript $\pm$. The list is non-exhaustive and
    reflects our current knowledge.
}\label{tab:Particles}

\begin{tabular}{llllllc}
\midrule
particle & notation & simplified notation & charge $Q$  & charge $P$  & composition & colour code \\
\midrule
head  & $\mathrm{h}^{\pm}$ & $\mathrm{h}^{\pm}$ & $\pm\frac{1}{2}$  & 0 & &\includegraphics[height=12pt]{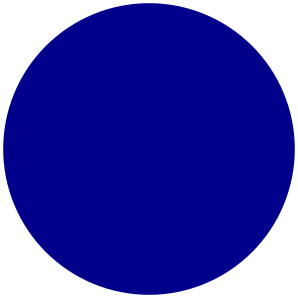}\\
tail   & $\mathrm{t}^{\pm}$ & $\mathrm{t}^{\pm}$ & $\pm\frac{1}{2}$& 0 &  &\includegraphics[height=12pt]{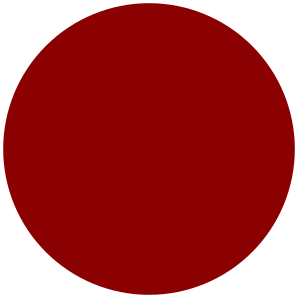}\\
pivot  & $\mathrm{p}_{\pm}$ & $\mathrm{p}^{\pm}$ & $0$  & $\pm\frac{1}{2}$ & &\includegraphics[height=12pt]{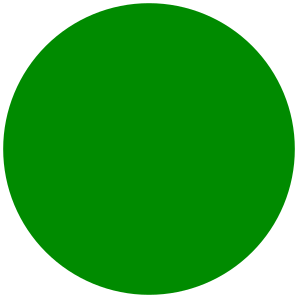}\\
\midrule
filament; tip & $\mathrm{f}^{\pm\pm}$   & $\mathrm{f}^{\pm}$ & $\pm 1$  & $0$ & h$^\pm$ + t$^\pm$ &\includegraphics[height=12pt]{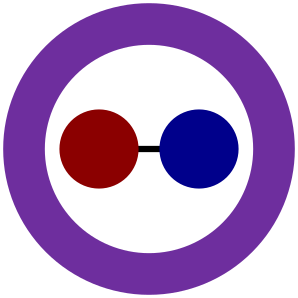}\\
core; phase singularity & $\mathrm{c}^{\pm\pm}_{\pm\pm}$   & $\mathrm{c}^{\pm}$ & $\pm 1$  & $\pm 1$ & h$^\pm$ + t$^\pm$ + 2p$_\pm$ & \includegraphics[height=12pt]{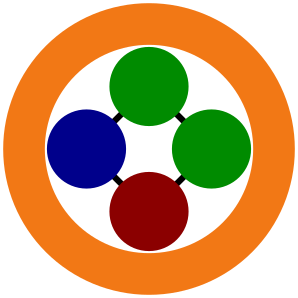}\\
shrink site & $\mathrm{s}^{\pm}_{\pm}$ & $\mathrm{s}^{\pm}$ & $\pm\frac{1}{2}$  & $\pm\frac{1}{2}$  & p$_\pm$ + t$^\pm$ &
\includegraphics[height=12pt]{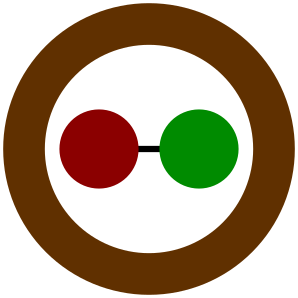}\\
growth site & $\mathrm{g}^{\pm\pm}_{\pm}$   & $\mathrm{g}^{\pm}$ & $\pm 1$  & $\pm\frac{1}{2}$ & h$^\pm$ + t$^\pm$ + p$_\pm$ & \includegraphics[height=12pt]{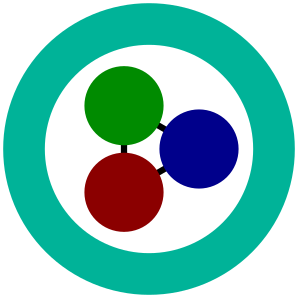}\\
\midrule
\end{tabular}
\end{center}
\end{table}

As we illustrate our general theoretical framework below with an analysis of experiments performed on cultured human heart cells \cite{Harlaar:2021}, we now describe the current state-of-art and associated challenges in the description and understanding of cardiac excitation patterns.

%Gaining a better understanding of the complex processes at different spatiotemporal scales during heart rhythm disorders is a major challenge.
The mechanical contraction of each myocyte in the heart is triggered by electrical depolarisation of the cells, which first propagates through a dedicated conduction system, but is thereafter passed on from cardiomyocyte to cardiomyocyte. As a result, the activation wave can re-excite itself, or a local focal source can start extra excitations. Such abnormal emerging behaviour manifests as a cardiac arrhythmia.

A simple arrhythmia in which the excitation wave follows a fixed path in time, or which originates from a region of ectopic impulse formation can be treated by ablation therapy, in which the trigger or path is physically destroyed \cite{haissaguerre_spontaneous_1998}. Even though efficiency has improved significantly throughout the years, the success rate is rather low for arrhythmias with incompletely understood spatiotemporal organisation \cite{cronin_2019,mujovic_catheter_2017}.
Such arrhythmias include ventricular fibrillation \cite{Gray:1998, Jalife:1998}, which is the most lethal one, and atrial fibrillation, which affects approximately \qty{1}{\percent} of the world population, and is responsible for one third of stroke cases \cite{Samol:2016}. In both atrial and ventricular fibrillation, as well as in ventricular tachycardia, rotating vortices of activation have been observed \cite{Allessie:1973}, named spiral waves, scroll waves, or cardiac rotors. While computer simulations of the heart \cite{Gray:1995} and initial observations \cite{Gray:1998} often yield long-lived stable rotors, complex patterns detected in real hearts commonly show a more complex regime that exhibits conduction blocks and transient rotors which do not even complete a full turn. There is a consensus that interacting wave fragments, also called wavelets \cite{moe_atrial_1959,lee_reconsidering_2020,Arno:2021} play a role in atrial and ventricular fibrillation. However, the precise mechanisms of onset, sustainment, and termination of complex arrhythmias remain to be further elucidated \cite{aras_ventricular_2017,shibata_mechanism_2022}.

We now review the concepts of phase and phase singularities that are used to describe excitable media. Winfree\cite{Winfree:1973} already noted that many biological processes are cyclic in nature, and therefore could be described naturally by a `phase'. Near the centre of a rotating spiral wave, phase singularities may occur \cite{winfree_singular_1983}, which form filament curves in three dimensions. Given two observed variables $V$ and $R$ that depend on the position $\vec{r}$ and time $t$ in a spatiotemporal pattern, it is common to call the polar angle in the state space spanned by $V$ and $R$ the `phase of activation':
%Gray~\emph{et~al.} calculated the spatiotemporal phase in a ventricular fibrillation pattern as \cite{Gray:1998}:
\begin{align}
    \varphi(\vec{r},t) &= \operatorname{atan2}( R(\vec{r},t) - R_*,  V(\vec{r},t) - V_*).  \label{phi_act}
\end{align}
Here, $\operatorname{atan2}$ is the two-argument arc-tangent function, delivering values in $(-\pi,\pi]$. In a seminal work, Gray~\emph{et~al.} analysed ventricular fibrillation patterns in this manner \cite{Gray:1998}, with $V(\vec{r},t)$ the optical intensity representing normalised transmembrane potential, and $R(\vec{r},t) = V(\vec{r},t+\tau_0)$ its time-delayed version.
The main idea here is that points that differ in phase by an integer multiple of $2\pi$ are effectively in the same state. After spatial filtering of the signal, distinct points occurred where all phases meet, called phase singularities \cite{winfree_singular_1983}. Since then, analysis tools \cite{Gray:1998,Bray:2003}, ablation strategies \cite{Narayan:2012}, and mathematical theory \cite{Clayton:2005,PanfilovDierckx:2017} have been developed based on the assumed existence of phase singularities during arrhythmias. A phase singularity can also be regarded as the intersection of two contour lines of two different state variables, which also allows to locate them numerically\cite{Fenton:1998}. An extension of this topological approach that focuses on the dynamical transitions between the contour intersections was described by Marcotte and Grigoriev \cite{Marcotte:2017}.
%If one visualizes the phase distribution $\varphi(x,y)$ on a portion of the heart as the height above the XY-plane, this phase surface close to a phase singularity looks like a staircase surface \cite{Tomii:2021,Arno:2021}. If the medium is activated multiple times, this `staircase' can have several windings, and the phase becomes a multivalued function. In the mathematical branch of complex analysis, the multivalued graph of phase is known as a Riemann surface. It can be reduced to a proper function by choosing one leaf of the surface, which is then bounded by a so-called branch cut, across which the phase difference is $\pm 2\pi$. Such branch cut is seen in isochrone maphase singularity as the border between regions that are first and last activated over time.

Characterising excitable media in terms of phase singularities was pioneered based on observations in oscillating chemical reactions \cite{zhabotinsky_autowave_1973,Winfree:1973}. When a phase singularity is tracked in computational models of cardiac excitation \cite{Fenton:2008}, however, it is often seen to follow a star-like or zig-zag pattern \cite{Krinsky:1992}. Such a `linear rotor core' also occurs at the interface between regions with different action potential duration, where it was historically called a `reverbator' \cite{krinsky_spread_1966,krinsky_fibrillation_1968}.
Linear core rotors have for decades escaped a thorough theoretical description, in contrast to the better-behaved `rigidly rotating rotors'. The latter are now well understood in terms of response function theory \cite{Keener:1988, Biktashev:1994,Henry:2000,Verschelde:2007}, which enables to forecast their motion if they are long-lived and weakly interacting with other rotors, boundaries or inhomogeneities. However, in arrhythmia management, the key processes to control are the transient processes causing the `birth' of rotors from wave breaks during arrhythmogenesis, and the elimination of rotors under pharmacological treatment, scarification of the medium, or by mutual collisions. Even worse, atrial and ventricular fibrillation are characterised by an incessant interaction between short-lived wavebreaks, for which a theory is hitherto largely lacking.

In recent years, it was suggested independently by Tomii~\emph{et~al.}\cite{Tomii:2021} and Arno~\emph{et~al.}\cite{Arno:2021} that the classical phase singularity concept may not be an accurate description for the structure of a linear rotor core. Here, the zig-zag pattern emerges since the classical spiral wave tip, i.e., the end point of the wave front, is moving along the border of a region that has not yet recovered from the previous activation. As soon as this region recovers, the spiral wave tip turns over about \qty{180}{\degree} in a so-called pivot point and then travels in the opposite direction, after which this process is repeated.
The mentioned authors \cite{Tomii:2021, Arno:2021} remarked that the parts of the medium at both sides of the linear rotor core have a different phase, as these regions were excited at different times. Hence, Tomii~\emph{et~al.} coined that across the conduction block line at the centre of a linear-core rotor, there is a phase discontinuity. We also noted this \cite{Arno:2021}, but remarked that due to electrotonic effects, i.e., diffusion of transmembrane potential, a boundary layer of finite width will be formed, across which the phase appears to change rapidly in space. The situation is reminiscent of domain walls or other interfaces in physics, and therefore we refer to them as `phase defects' \cite{Arno:2021}. The diffusive effect also explains why the phase defects were not identified before: If you have a line along which phase is discontinuous and smoothen the variables slightly, the resulting phase function tends to become continuous except in isolated points that happen to have the values $V=V_*$, $R=R_*$, where all phases meet. These points are exactly the phase singularities that have been used in cardiac analysis for nearly three decades now \cite{Gray:1995}. While many of these points will be true phase singularities, a diffusive smoothing applied to a conduction block may introduce pairs of false phase singularities, which in our opinion explains the non-robust detection of phase singularity pairs in conduction block regions \cite{Arno:2021, Rodrigo:2017}.

The concept of a `phase defect' in excitable media is able to link different objects already known in the field and will reveal new ones, as can be seen below. From the explanation above, the phase defect can represent a conduction block without a rotor attached to it, and the conduction block that is present in a linear-core rotor. In hindsight, Krinsky's reverbators \cite{krinsky_spread_1966,krinsky_fibrillation_1968} were also phase defects, as across the interface, the phase was also discontinuous. Note that conduction blocks are key features in activation maps \cite{Janse:1980}. In clinical acquisitions, it is common to record the local activation times (LATs) and draw the isochrones. Where different isochrones coincide, a conduction block and hence a phase defect is present. This situation is similar to cartography, where coincident isolines mark a cliff, i.e., a discontinuity in altitude.

The aim of this paper is to provide a more detailed theoretical framework that reconciles the concepts of wave fronts, wave backs, conduction blocks, rotors with different core types, fragmented wave fronts (wavelets) etc. The key concept will be the phase defect mentioned above, and we will show that tracking how such phase defect is connected to wave fronts and backs opens up new ways for pattern analysis. Our exposition has been designed to unravel the critical processes that cause, sustain or terminate complex arrhythmias. We purposely use terminology from particle physics, as in this work we bring several analogies from that discipline to a cardiac context.

\section*{Conceptual framework}\label{sec:Concepts}

\begin{figure*}[t]
    \centering
    \includegraphics[width=0.7\textwidth]{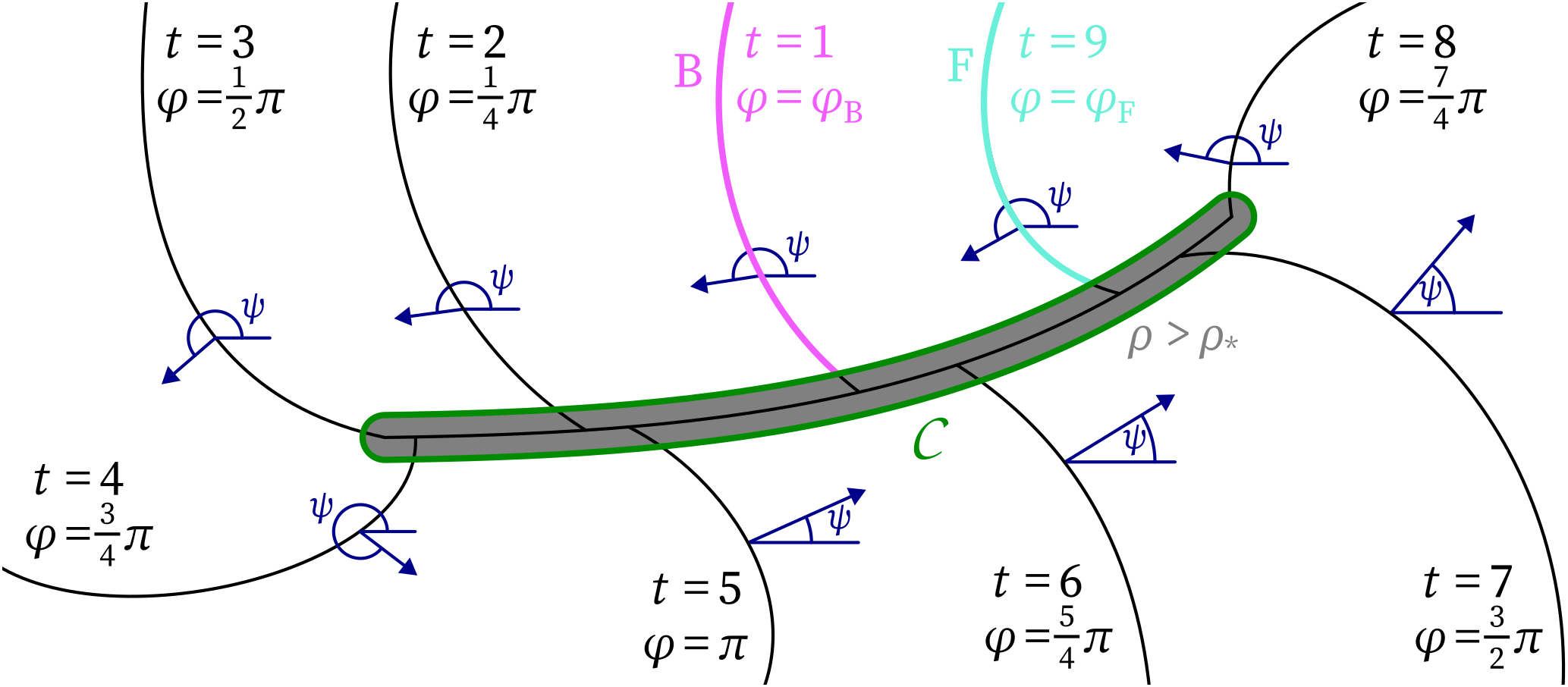}
    \caption[
        Sketch illustrating the main quantities in the definitions for the topological framework.
    ]{
        Sketch illustrating the main quantities in the definitions for the topological framework on a linear-core rotor.
        Isochrones at different LATs $t$ are used to define the phase $\varphi$,
        which in turn is used to define the wave front ($\mathrm{F}$) and back ($\mathrm{B}$).
        At the core of the vertex, the phase $\varphi$ is discontinuous.
        The threshold $\rho > \rho_*$ of the phase defect density is used to define the phase defect region $Z$.
        The angle $\psi$ with the $x$-axis is the direction of wave propagation.
    }
    \label{fig:concepts}
\end{figure*}

We build up the framework by defining and motivating subsequently: phase, phase defects, quasiparticles, charges and diagrams.
A sketch illustrating the main quantities in the following definitions can be found in Fig.~\ref{fig:concepts}.

\subsection*{Phase}
Winfree wrote the following motivation to describe biological media with a phase concept\cite{winfree_singular_1983}: ``After excitation, [the phase] initially grows linearly with time, but after a while, it lingers near a value of $2\pi$, which is equivalent to phase 0, the completely recovered state.'' So, when looking at an action potential as a function of time, we require that $\varphi(t) = 0$ for $t<0$ and $\varphi(t) = 2\pi$ for large $t$. Phase definition \eqref{phi_act} is one possibility. It is certainly not unique, since different variables $V$, $R$ can be chosen, as well as their threshold values $V_*$, $R_*$. Two other choices were given in by Kabus~\emph{et~al.}\cite{kabus2022numerical}, designed to exhibit no sharp phase variation around the wave front.

To make the link with isochrone maps, we here define the local activation time (LAT) as the last time when the transmembrane potential $V$ rose above a threshold value $V_*$:
\begin{align}
    t_\mathrm{LAT}(\vec{r},t) = \operatorname{argmax}_{t_0 < t} \{
        V(\vec{r},t_0) = V_*
        \wedge
        \partial_t V(\vec{r},t_0) > 0
    \}
\end{align}

The elapsed time since last activation is then $\tau = t - t_\mathrm{LAT}$. We now propose a linear activation time-based phase:
\begin{align}
    \varphi(\vec{r}, t) =
    \begin{cases}
        2\pi \frac{\tau(\vec{r}, t)}{\tau_0} & \tau(\vec{r}, t) < \tau_0 \\
        2 \pi & \tau(\vec{r}, t) \geq \tau_0
    \end{cases}
    \label{phi_lin}
\end{align}
Hence, the resting state has phase 0, and after activation phase grows linearly with time over a time interval $\tau_0$, after which the tissue is assumed to be fully recovered. Then the tissue has phase $2\pi$, which is equivalent to phase $0$. This simple phase has its limitations: It is identical to zero at the foot of the upstroke, when $V$ is still lower than $V_*$. After a time interval $\tau_0$, that we choose equal to the local action potential duration in that point, the medium is assumed to be fully recovered. Still, it suffices for the formal introduction of the concepts below.

\subsection*{Phase defects}
By construction, $\varphi$ is continuous over a wave front, since there the phase amounts to $0$. It is also continuous over a wave back, where $\varphi$ is slightly lower than $2\pi$. At conduction block sites, however, the impeding wave does not make $V$ cross the threshold value $V_*$ from below anymore, and hence $\varphi$ will be discontinuous across a conduction block. As a consequence of the phase definition \eqref{phi_lin}, a true phase discontinuity forms, of co-dimension 1. In an inhomogeneous medium where action potential duration depends on position, the edge of the inhomogeneity can take any shape, such that the discontinuity of $\varphi$ needs not be a line but can take any shape: Closed or branching phase defect structures may also form.

To numerically detect the phase defect, we compute the phase defect density $\rho(\vec{r},t)$ over all pixels of the grid
%\cite{kabus2022numerical,Kuramoto:1984}:
%\begin{align}
%    \rho(\vec{r}_a,t) = 1 - \frac{1}{N_a} \left| \sum_{b \in \mathcal{N}(a)} e^{i \varphi(\vec{r}_b,t)} \right| \label{rho}
%\end{align}
\cite{kabus2022numerical,Tomii:2021}:
\begin{align}
    \rho(\vec{r}_a,t)
    =
    \frac{1}{2N_a}
    \sum_{b \in \mathcal{N}(a)}
    [1 - \cos(
        \varphi(\vec{r}_a,t)
        -
        \varphi(\vec{r}_b,t)
    )]
    \label{rho}
\end{align}
Here, $\mathcal{N}(a)$ is the set of nearest neighbours in the Cartesian grid next to the point $\vec{r}_a$. $N_a$ is the number of such neighbours; it is everywhere four except near the domain boundaries or inexcitable obstacles. When the phases locally differ much, $\rho$ will have a value close to $1$. If all excitable elements excite simultaneously, $\rho=0$. The set $\mathrm{Z}$ of points inside a phase defect can be identified numerically as the set where $\rho$ exceeds a chosen threshold value $\rho_*$ \cite{kabus2022numerical}:
\begin{align}
    \mathrm{Z}(t) =  \{
        \vec{r}
        \;|\;
        \rho(\vec r, t) > \rho_*
    \}
    \label{Z}
\end{align}
In this manner, the phase defect region $\mathrm{Z}$ forms an open subset of the domain; it is not a one-dimensional curve anymore. The contours $\mathcal{C}$ obeying $\rho = \rho_*$ encircling each phase defect are well-defined and will be used below.

It is also possible to start from the classical activation phase, rescale it via a sigmoid function and then compute the phase defect density\cite{kabus2022numerical}. In that case, the phase defect also becomes an open set $\mathrm{Z}$, surrounded by a continuous contour $\mathcal{C}$.

\subsection*{Special points}

The essence of our analysis is to divide the medium in three regions and to identify the points where all these regions meet, see Fig.~\ref{fig:intro}.

First, we compute the phase defect density $\rho$ via \eqref{rho} to find all points within the phase defect $\mathrm{Z}$, see \eqref{Z}. These points will be colored in gray.

Let $0<\varphi_\mathrm{F}<\varphi_\mathrm{B}<2\pi$,
where $\varphi_\mathrm{F}$ is a small phase threshold close to $0$,
and $\varphi_\mathrm{B}$ a large phase threshold close to $2\pi$, which must be small enough that a jump from $\varphi_\mathrm{B}$ to $\varphi_\mathrm{F}$ is considered a phase defect following \eqref{rho} and \eqref{Z}.
The set of points where the phase is between these values and outside a phase defect is the excited region ($\mathrm{E}$):
\begin{align}
\mathrm{E}(t) =  \{
    \vec{r}\;|\;
    \rho(\vec r, t) < \rho_*
    \wedge
    \varphi(\vec r, t) \in (\varphi_\mathrm{F}, \varphi_\mathrm{B})
\}
\end{align}
and is depicted in yellow in Fig.~\ref{fig:intro}.
The set of points outside of that range and phase defects is called unexcited ($\mathrm{U}$) and coloured white:
\begin{align}
\mathrm{U}(t) =  \{
    \vec{r}\;|\;
    \rho(\vec r, t) < \rho_*
    \wedge
    \varphi(\vec r, t) \not\in [\varphi_\mathrm{F}, \varphi_\mathrm{B}]
\}
\end{align}

Where the excited region meets the unexcited one, the wave front ($\mathrm{F}$) and wave back ($\mathrm{B}$) are found, denoted in cyan and magenta, respectively:
\begin{align}
\mathrm{F}(t) =  \{
    \vec{r}\;|\;
    \rho(\vec r, t) < \rho_*
    \wedge
    \varphi(\vec r, t) = \varphi_\mathrm{F}
\}
\nonumber \\
\mathrm{B}(t) =  \{
    \vec{r}\;|\;
    \rho(\vec r, t) < \rho_*
    \wedge
    \varphi(\vec r, t) = \varphi_\mathrm{B}
\}
\end{align}

For completeness, we remark that two other boundaries of co-dimension 1 exist: the border between $\mathrm{Z}$ and $\mathrm{U}$, and the border between $\mathrm{Z}$ and $\mathrm{E}$. Together they form the aforementioned contours $\mathcal{C}$.

In the classical theory of excitation,
the wave front meets the wave back in a single point, and this point coincides with the classical phase singularity \cite{Gray:1995,Zykov:1987,Clayton:2005}.

In our case, wave fronts and wave backs will always end on a phase defects and medium boundaries. We call the point where a wave front and a phase defect meet, a `head' (denoted in blue) and a point where a wave back and a phase defect meet, a `tail' (denoted in red). Essentially, heads and tails are the loci where the three distinct zones ($\mathrm{E}$, $\mathrm{U}$ and $\mathrm{Z}$) meet. As such, any continuous deformation of the boundaries between these zones will not affect the existence of such special points. Therefore, we call heads and tails topologically protected. A similar reasoning holds under time evolution: If the excited, unexcited and phase defect regions change their boundaries in a continuous manner, the head point can move in time, but is only lost if it meets another head or a medium boundary. An analogous reasoning justifies to call the tail points to be topologically conserved.

From Fig.~\ref{fig:intro}E, we see that heads can exist with two different chiralities: If one circumscribes the head point counterclockwise and meets the regions $\mathrm{Z}$, $\mathrm{E}$, $\mathrm{U}$ in that order, we call it a `positive' head, denoted $\mathrm{h}^+$. If we find the order $\mathrm{Z}$, $\mathrm{U}$, $\mathrm{E}$, we call it a `negative' head, written $\mathrm{h}^-$. These intuitive notions will be formalized below, to grant them the meaning of a `charge'.

For now, let us return to the depiction of a linear-core rotor in Fig.~\ref{fig:intro}D. We observe yet another type of special point, namely the end point of the phase defect line. Such point are already known in literature as the loci around which the wave turns suddenly, and hence we also call them `pivots'. Using the LAT-based phase definition \eqref{phi_lin}, it follows that the phase defects are truly curves of discontinuous phase. Hence these lines can have distinct end points, the pivots. The phase defect lines may also branch spatially; in such case a phase defect `joint' will emerge, which however falls outside the scope of our present exposition. Pivots also exist in two flavours, see Fig.~\ref{fig:intro}E:
If the wave made a sudden half-turn to the left at that pivot, we call the pivot point a $\mathrm{p}_+$, with the plus sign marking counterclockwise rotation.
If the half-turn was to the right, we call it a $\mathrm{p}_-$.
We will discuss the topological preservation of pivots below, after the formal introduction of pivot charge.

\subsection*{Two topological charges: rotation in state space vs. rotation in physical space}

We will here demonstrate that heads and tails carry half-integer charges that are identical to the classical `charge' of a phase singularity. Pivots carry a charge that is also half-integer, but of a different kind.

First, consider a portion of the contour $\mathcal{C}$ that surrounds a phase defect in a counterclockwise sense, i.e., with the phase defect to the left. See Fig. \ref{fig:concepts}. With arc length parameter $\ell$, we evaluate the classical charge density:
\begin{align}
    Q
    =
    - \frac{1}{2\pi} \int
    d \varphi
    =
    - \frac{1}{2\pi} \int
    \vec{\nabla} \varphi \cdot \vec{d \ell}. \label{intQ}
\end{align}
Here, $\varphi$ can be any phase related to the local state of cells; both \eqref{phi_act} and \eqref{phi_lin} are possible, just like other choices\cite{Gray:1995,Clayton:2005,Tomii:2021}. The minus sign is included such that a counterclockwise rotating spiral has $Q=+1$. Authors usually evaluate \eqref{intQ} around a closed contour, to find that $Q$ always takes an integer value if $\varphi$ is well-defined over the entire contour.
However, if one takes only a portion of the contour, one can find the charge density $dQ = - \frac{d\varphi}{2\pi}$. Part of the phase change occurs during excitation, and part during the biological recovery process. If one assumes these changes to take place at the wave front and wave back, a phase change of $\pm \pi$ will be found each time the contour encircling the phase defect meets a wave front or wave back. These values correspond to localized charges of $\pm \frac{1}{2}$. Therefore, we here choose to ascribe a topological charge of $\pm \frac{1}{2}$ to heads and tails, with the sign chosen as in Fig.~\ref{fig:intro}E. In principle, this choice can be made rigorous by defining the phase in the region $\mathrm{U}$ to be 0 and the phase over the region $\mathrm{E}$ to be $\pi$. Elegantly, the sum of all head and tail charges when circumscribing a phase defect equals the classical topological charge $Q$. To distinguish it from the pivot charge below, we call this the $Q$-charge. From the above, it follows that the $Q$-charge in a way `measures' the amount of rotation in state space.

Second, we introduce the pivot charge as follows. Define the angle of wave propagation as:
\begin{align}
    \psi = \operatorname{atan2}( \partial_y t_\mathrm{LAT}, \partial_x t_\mathrm{LAT}).
\end{align}
where the LAT is continuous and differentiable. $\psi$ is the angle with the positive $x$-direction under which the last arrived activation wave travelled. Now, in full analogy with \eqref{intQ}, step along the contour $\mathcal{C}$ while keeping the phase defect to the left:
\begin{align}
    P
    =
    \frac{1}{2\pi}
    \int
    d \psi
    =
    \int
    \vec{\nabla}\psi \cdot \vec{d \ell}
    \label{intP}
\end{align}

Let us verify how the $P$-charge density $d P = \frac{d\psi}{2\pi}$ behaves in standard cases. First, take the neighbourhood of the positive pivot at the top-right panel of Fig.~\ref{fig:intro}E. There, the wave turns over $+\pi$ near the pivot, such that $P=+\frac{1}{2}$. This corresponds to the intuitive notion of `half a turn to the left' that we advocated above. Similarly, a negative pivot $\mathrm{p}_-$ can now be proven to have $P=-\frac{1}{2}$. Third, consider a classical spiral wave core denoted as $\mathrm{c}^+$ in the bottom row of Fig.~\ref{fig:intro}E. Near the classical phase singularity, phase gradients will become arbitrarily large, such that the classical phase singularity is also a special case of a phase defect: It is a phase defect of small size\cite{Arno:2021}, whose exact diameter depends on the chosen $\rho_*$. Evaluating $P$ around a closed contour that circumscribes the phase singularity counterclockwise delivers: $P=+1$. Noteworthily, even circular rotor cores form over time from conduction blocks: Initially, a finite conduction block line is created, e.g. by an S1-S2 stimulus protocol, with two pivots each having $P=+\frac{1}{2}$. During successive wave rotations, this phase defect shrinks to a point. At that time, the two pivots with $P=+\frac{1}{2}$ as well as the head and tail with each $Q=+\frac{1}{2}$ merge to form a stable state, with $P=+1$ and $Q=+1$, that we call a `core particle' here.

Let us now investigate when the $P$-charge is conserved. Hereto, we apply Stokes' theorem to \eqref{intP} on a closed contour $\mathcal{C}$ around a region $S$.
\begin{align}
    P = \frac{1}{2\pi} \oint d \psi =\oint \vec{\nabla}\psi \cdot \vec{d \ell}  =
    \iint \vec{\nabla} \times \vec{\nabla}\psi \cdot \vec{d S} =0
    \label{stokes}
\end{align}
The last equality follows from the properties of curl and gradient. The second-last equality requires that the conditions for Stokes' theorem are fulfilled: That $\vec{\nabla}\psi$ is everywhere in $S$ defined, and that it has continuous first derivatives. Result \eqref{stokes} implies that $P$ will not change its value if the contour is continuously deformed outside regions with phase defects or phase singularities (where $\psi$ is undefined), or where $\psi$ suddenly changes its orientation. The latter case happens at excitation sources, or on the boundary at the position where it was hit by a wave front. Given these limitations, the $P$-charge is conserved. E.g. a closed contour may contain two opposite rotating linear-core rotors, i.e., with $2\mathrm{p}_+$ and $2\mathrm{p}_-$ in it. The total $P=\frac{1}{2}+\frac{1}{2}-\frac{1}{2}-\frac{1}{2}=0$. If these rotors merge to form a single conduction block line, a $\mathrm{p}_+$, $\mathrm{p}_-$ pair annihilates, leaving the second $\mathrm{p}_+, \mathrm{p}_-$ inside the contour, and still $P=\frac{1}{2}-\frac{1}{2}=0$.

In conclusion, the $P$-charge measures the amount of rotation in physical space. It is thus different from the $Q$-charge, carried by heads and tails. Therefore, when using the detailed notation for charges (see Tab. \ref{tab:Particles}), we denote the $P$-charge in subscript: $\mathrm{p}_\pm$.

\subsection*{Quasiparticles}

Until here, we have identified three different special points in excitable media, that can carry half-integer `charges' and that seem to be preserved under time evolution (for now disregarding boundaries and obstacles). We propose to refer to these special points as quasiparticles and call them `cardions'. Examples of the seven different cardions that we know now of are given in Fig.~\ref{fig:intro}E. While three cardions are elementary, heads, tails and pivots, four of them arise as bound states between the elementary ones.

\subsection*{Bound states}

Strikingly, our analysis of activation patterns in simulation and experiments revealed that the heads, tails and pivots tend to form bound states, see Fig.~\ref{fig:intro}E and Table \ref{tab:Particles}. The situation reminds us of the hydrogen atom consisting of a proton and an electron, or the proton itself, which is made up of quarks in particle physics.

The first bound state consists of a head and a tail of equal $Q$-charge. Already in Fig.~\ref{fig:intro}D, it can be seen that while the tail traces a phase defect, a head can follow it directly, exciting the newly-recovered tissue created by the tail particle. The two special points propagate as a pair along the phase defect, with an excitable gap between them. In the classical viewpoint, this corresponds to a wave front and wave back ending in one point, the classical `spiral tip'. In three spatial dimensions, the tips form a filament curve \cite{Winfree:1996,Clayton:2005} around which the wave rotates.
Therefore, we annotate the bound state of head and tail with f and call it a filament point and annotate it in the colour purple. The convention of assigning half a $Q$-charge to heads and tails has the useful property that the sum of these charges will be equal to the classical topological charge $Q=\frac{1}{2}+ \frac{1}{2} = 1$ of the filament. As the absolute charge of a filament particle is always the same, we abbreviate the `full' notation f$^{\pm \pm}$ to $\mathrm{f}^{\pm}$, see second column in Tab. \ref{tab:Particles}. In this simpler notation, we write the charges of all particles with superscript, for example, $\mathrm{p}^+$, rather than $\mathrm{p}_+$.

The second bound state is the core particle, which we introduced above as an example of $P$-charge. This core particle occurs with rigidly rotating spiral waves, where the phase defect is small and head and tails are close to each other. Since these cores typically form from shrinking of a phase defect line containing two pivots of the same charge, the core particle consists of a head, a tail and two pivots (where the latter are no more distinguishable). This `core particle', denoted $\mathrm{c}^{\pm\pm}_{\pm\pm}$ or shorter $\mathrm{c}^{\pm}$ and indicated by an orange circle, see the bottom panel of Fig.~\ref{fig:intro}E.

The next two bound states have no classical counterpart to our knowledge. Consider a propagating wave front that hits a wave back. Such event will generally occur in a single point first, creating a small phase defect. The end point of one side of the wave front will then travel along the existing wave back, and in the process, the conduction block region grows to become a phase defect line. Hence, there exists a state of a head, tail, and pivot of the same chirality that all move together. We call this a `growth' particle and denote it with a teal circle. It has $Q=\pm 1$ and $P=\pm \frac{1}{2}$, with the same sign chosen. Hence, we identify quasiparticles g$^{++}_+$ and $\mathrm{g}^{--}_{-}$, or $\mathrm{g}^+$, $\mathrm{g}^-$ in shortened notation (see Tab. \ref{tab:Particles}).

Finally, we were ourselves struck by the observation that the shrinking and growing of phase defect lines occurs in an asymmetric manner. When a wave back travels along a phase defect such that the tissue has recovered at both sides of the defect line, the phase defect line will retract, see Fig.~\ref{fig:intro}E, bottom right. Such retraction implies that a tail and a pivot of same chirality travel together. We denote this as a `shrink' particle s$^\pm_\pm$, shortened s$^\pm$, and mark it in brown. Tab. \ref{tab:Particles} summarises the hitherto discovered quasiparticles and their bound states in excitable media.

\subsection*{Medium boundaries and conservation laws}

Until here, we have disregarded inexcitable obstacles in the medium, which also includes the medium boundary itself.

Let us first consider how the $Q$-charge behaves in the presence of obstacles. In the example of Fig.~\ref{fig:intro}D, the total $Q$-charge in the medium disregarding the medium boundaries would be $-\frac{1}{2}-\frac{1}{2} = -1$. However, we now argue that it might be useful to also regard medium boundaries as a kind of phase defect. If we do this, e.g. by colouring the medium boundaries also in grey, it follows that every wave front is either closed or has two endpoints (heads) with different chirality.
A similar reasoning holds for the wave backs, which are either closed, or end in two tails of opposite charge.
Then, by construction the total $Q$-charge for both all heads in $H$, the set of heads, and tails in $T$, the set of tails, must be zero:
\begin{align}
\sum_{h \in H}  Q_h &= 0
\\
\sum_{t \in T}  Q_t &= 0
\end{align}
These relations only hold if medium boundaries are considered as phase defects.

Taking the sum of the previous two statements proves that the total $Q$-charge in a two-dimensional medium is also always zero:
\begin{align}
\sum_{j \in H \cup T}  Q_j &= 0
\end{align}
considering medium boundaries as phase defects.
This result is stronger than the conservation of phase singularities, which only holds in the bulk of the medium.

In the viewpoint that includes medium boundaries and inexcitable obstacles, the classical annihilation of a rotor core at the medium boundary can be regarded as the head-tail pair becoming a pair of a head and tail which are travelling apart from each other on the boundary.  Also, the generation of a single rotor via S1-S2 stimulation now includes the boundary: The S2 pulse introduces a pair of heads, one of which joins a tail particle of the S1 pulse to form a filament quasiparticle.

An additional reason for including the medium boundary in the description lies in the interpretation of electrical signals. In leading order, the unipolar electrical potential measured by an electrode is determined by the angle subtended by the wave fronts and wave backs, seen from the electrode \cite{Holland:1977}. The measured signal is thus only determined by the boundaries of the wave fronts and backs, namely the heads and tail curves in three dimensions \cite{Arno:2024b}. Therefore, the concepts of heads and tails may also prove useful when improving efficient inverse methods to recover spatiotemporal activation sequences from patient measurements \cite{ramanathan_noninvasive_2004}.

\begin{figure*}[t]
    \centering
    \includegraphics[width=0.7\textwidth]{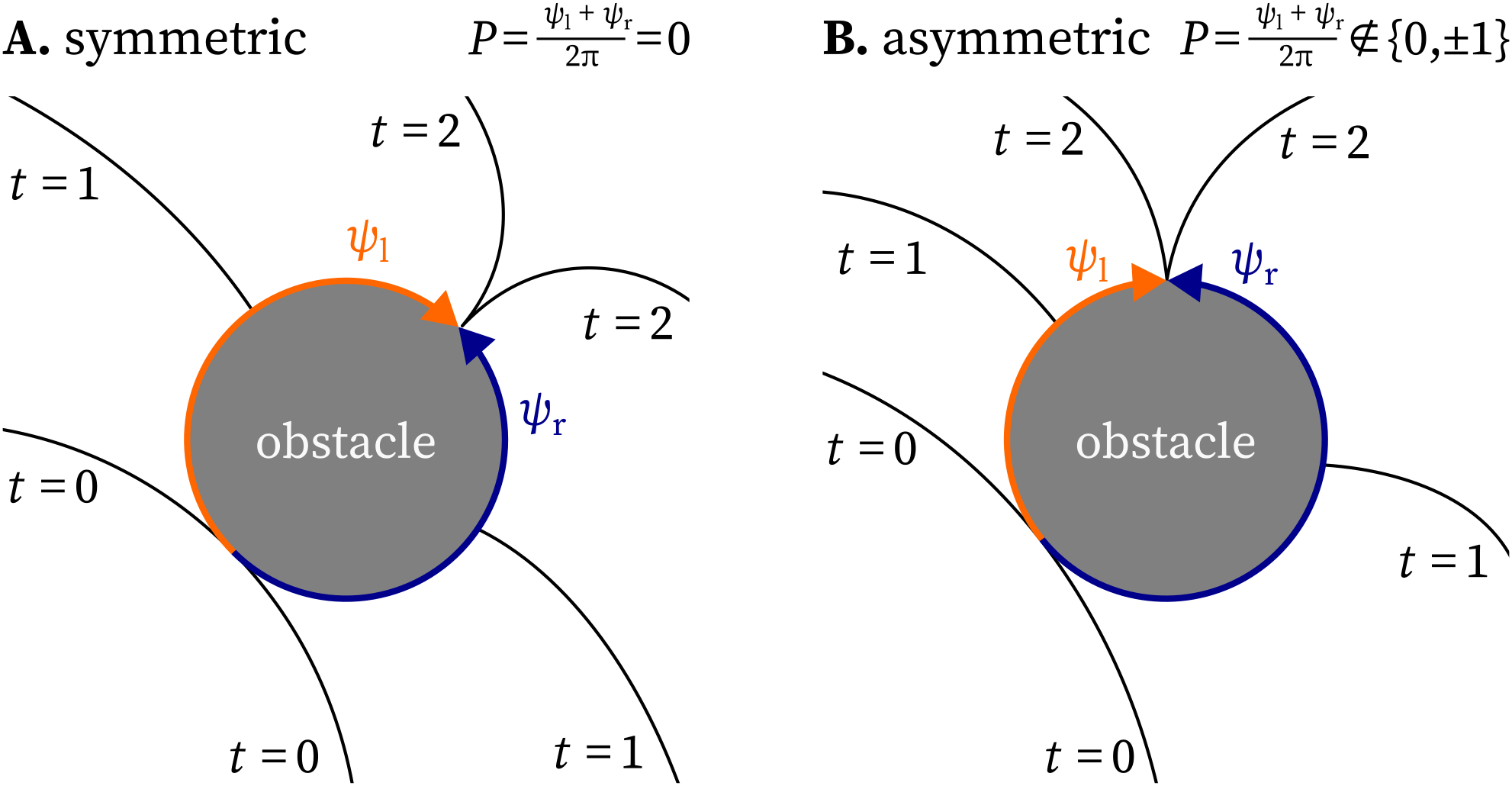}
    \caption{
        Non-integer $P$-charges are possible for paths around inexcitable obstacles.
    }
    \label{fig:obstacle:p}
\end{figure*}

Second, we consider how the pivot charge behaves in the presence of obstacles. If a wave front travels clockwise along the edge of a square domain, it will turn \qty{90}{\degree} to the right when crossing the corner. Thus, according to \eqref{intP}, the corners of a square domain carry a pivot charge of $\pm\frac{1}{4}$. Since the edge of obstacles can turn over any angle, pivot charges are not always half-integer, but can take on the boundary of the medium arbitrary values in $(-1,1)$. Another example is a wave hitting an obstacle within the medium, as shown in Fig.~\ref{fig:obstacle:p}. Suppose the wave encircling the obstacle counterclockwise travels over an angle $\psi_\mathrm{l} >0 $, while the wave travelling in the other direction around the obstacle over an angle $\psi_\mathrm{r} < 0$, with $|\psi_\mathrm{l}| + |\psi_\mathrm{r}| = 2\pi$ . Then, the $P$-charge of this obstacle will become via \eqref{intP}: $P = \frac{\psi_\mathrm{l} + \psi_\mathrm{r}}{2\pi} \in [-1, 1]$. In this situation, the $P$-charge is not conserved. Nonetheless, the value of the integral \eqref{intP} contains useful information: If it has $P = \pm 1$, a rotor is attached to the obstacle, while $P = 0$ indicates a symmetric conduction block. Intermediate values of $P$ indicate whether the wave-obstacle interaction is close to the formation of a rotor attached to the obstacle or not.

In summary, $P$-charge is not conserved in the presence of medium boundaries. This was also the case for the $Q$-charge with classical phase singularities, as they can simply vanish by collision with the medium boundary.

% the $Q$-charge is easily extended to include medium boundaries and obstacles, this is more difficult for the $P$-charge.  Within this work, we only associate pivot charges to line-shaped conduction blocks and do not consider their interaction with obstacles or medium boundaries.

\subsection*{Graphical depiction of processes via Feynman-like diagrams}

In the last step of our conceptual framework, we describe the evolution of cardions via diagrams similar to Feynman diagrams \cite{feynman_theory_1949}. A first example is given in Fig.~\ref{fig:intro}F. A $\mathrm{g}^+$, $\mathrm{g}^-$ pair is born, after which the growth particles each split into triples of head, tail, and pivot. In this graphical depiction, time passes from left to right and positions in space are represented along a single, vertical axis. The arrows represent the continuous existence of the quasiparticle in space and time, until an event happens in which the particle merges, splits or annihilates.
The premise of this work is that these diagrams can offer insight in the pattern's dynamics, see the examples below. % and we see potential on the longer term to use
%Within theoretical physics, such diagrams not only offer insight, but also serve as a guideline for advanced calculations and predictions of the system's behavior.

\section*{Results}

\subsection*{Figure of eight formation \emph{in silico}}

\begin{figure*}[tp]
    \centering
    \includegraphics[width=\textwidth]{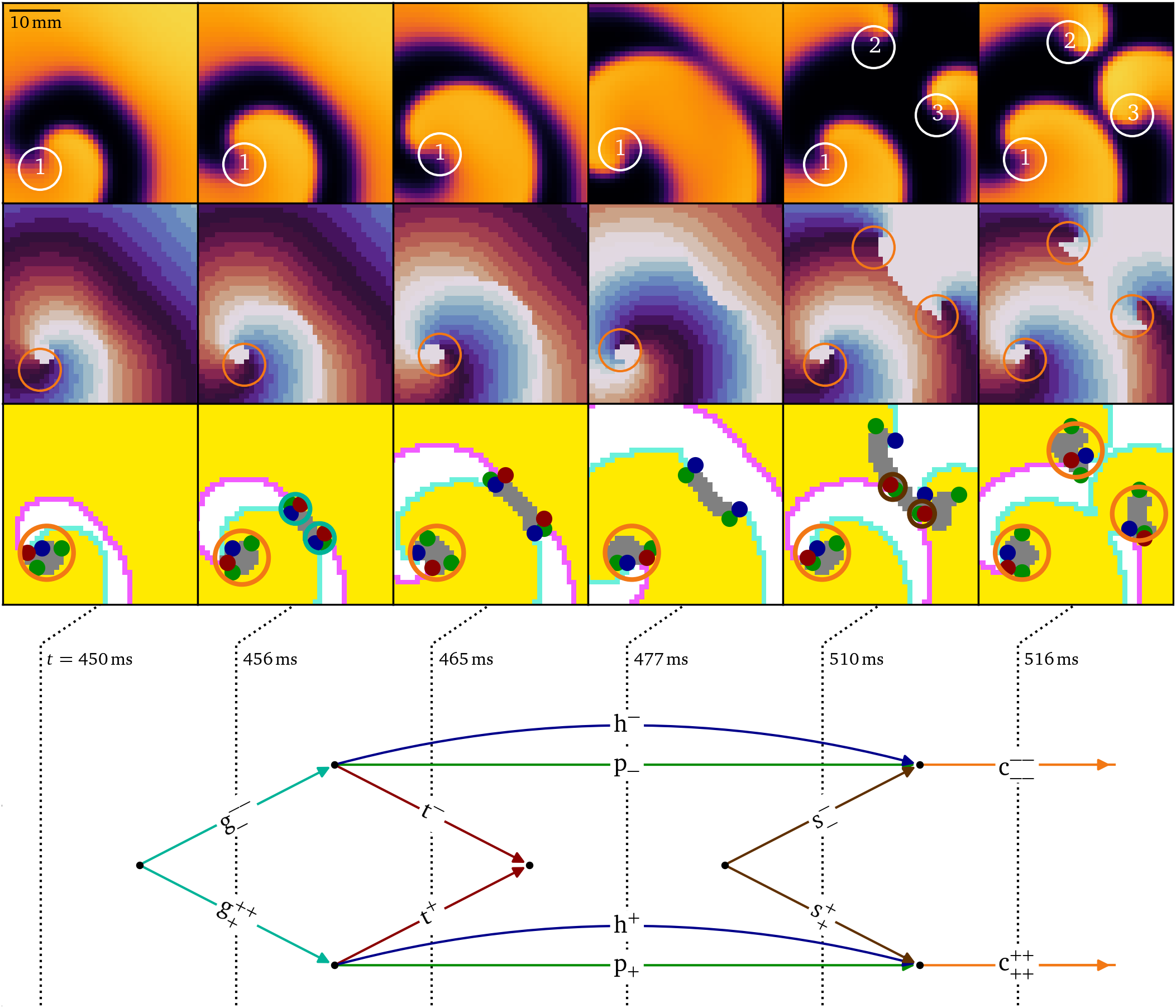}
    \caption[
        Diagrammatic analysis of the creation of two spiral waves in a simulation of break-up.
    ]{
        Diagrammatic analysis of the creation of two spiral waves in a simulation of break-up \cite{Karma:1993, Karma:1994}.
        First row: Snapshots of the normalized transmembrane potential
        with the phase singularities highlighted which were detected by the Kuklik method \cite{kuklik2017identification}, and
        in the second row, corresponding phase maps \cite{Gray:1998,kabus2022numerical}.
        Third row: Identification of the quasiparticles in subsequent snapshots.
        The resulting diagram at the bottom shows that the birth of a rotor pair involves seven topological interactions, including the creation of a shrink pair, and the splitting of a conduction block line.
        No heads, tails, or pivots are drawn at the boundary since it is an inset, not the medium boundary.
        The vortex with label 1 is excluded from the diagram of topological interactions.
        The same color scheme as in Fig.~\ref{fig:intro} is used consistently in this article.
    }
    \label{fig:smooka_creation}
\end{figure*}

Fig.~\ref{fig:smooka_creation} shows a numerical simulation in the smoothed Karma model \cite{Karma:1993,Marcotte:2017} for atrial fibrillation, in which a single rotor spontaneously breaks up into multiple vortices.
We analyze the creation of a vortex pair, known in cardiology literature as a figure-of-eight re-entry \cite{wu_double_1994}.
In the top two rows, phase singularities detected by the Kuklik method \cite{kuklik2017identification} are annotated on top of the normalised transmembrane voltage and phase.
The vortex pair with labels 2 and 3 is situated away from a central persisting rotor, labelled 1.
In the third row of panels, heads, tails and pivots, as well as their bound states are annotated.
When the wave front hits the wave back, a pair of $\mathrm{g}^+$, $\mathrm{g}^-$ emerges.
Each particle lasts until the vortex makes a pivoting turn around the formed conduction block line, which marks the decay of the $\mathrm{g}$-particles into triples of head, tail, and pivot. Just after $t=\qty{465}{\milli\second}$, the tail pair annihilates, as the wave back detaches from the conduction block line to form a single wave back again. The recovery of the tissue at the location of the initial collision marks the splitting of the phase defect line, which involves the birth of a shrink pair, just before $t=\qty{510}{\milli\second}$. Thereafter, both phase defect lines shrink until the head, tail, and two pivots are so close to each other that they effectively form a core particle. We conclude that the birth of a rotor pair, which is a single event in the classical theory, in fact consists of seven subsequent events, revealed by the vertices in the diagram
at the bottom of Fig.~\ref{fig:smooka_creation}.

\subsection*{Figure of eight creation \emph{in vitro} under burst pacing}

\begin{figure*}[t]
    \centering
    \includegraphics{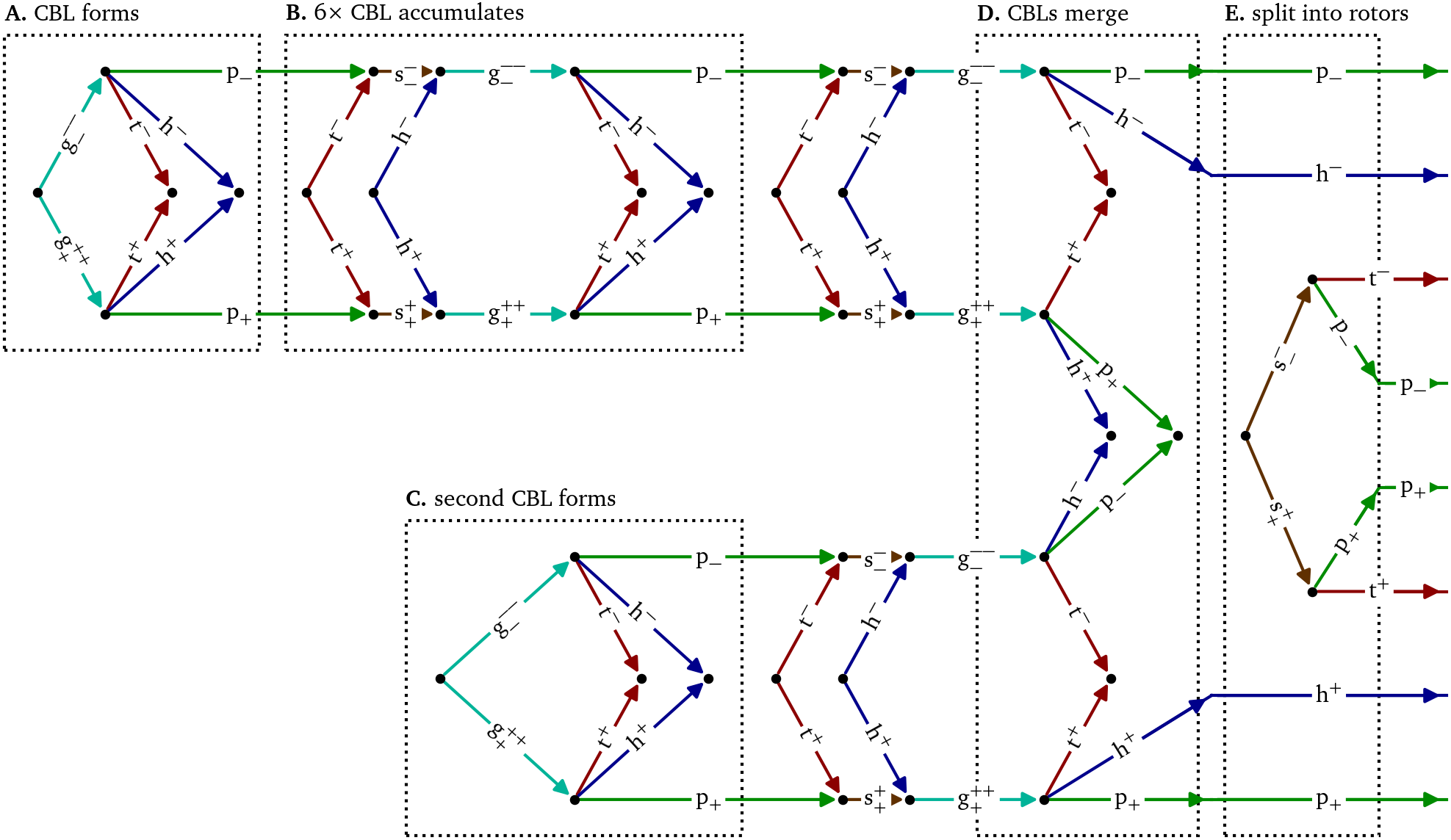}
    \caption[
        Feynman-like diagram of the creation of two spiral waves in an optical voltage mapping experiment.
    ]{
        Feynman-like diagram of the creation of two spiral waves via two merging conduction block lines (CBLs) in an optical voltage mapping experiment.
        The detailed quasiparticle viewpoint shows that both events are part of a single complex multi-stage process of up to twelve quasiparticles at the same time interacting in 93 reactions.
        The stages \textbf{A,~B,~D,~\&~E} are shown in more detail in Figs.~\ref{fig:hiam:init},~\ref{fig:hiam:accumulation},~\ref{fig:hiam:merge},~\&~\ref{fig:hiam:split}, respectively.
        The formation of the second CBL in panel~\textbf{C} is analogous to the first one in panel~\textbf{A} and Fig.~\ref{fig:hiam:init}.
    }
    \label{fig:hiam}
\end{figure*}

\begin{figure*}[tp]
    \centering
    \includegraphics[width=\textwidth]{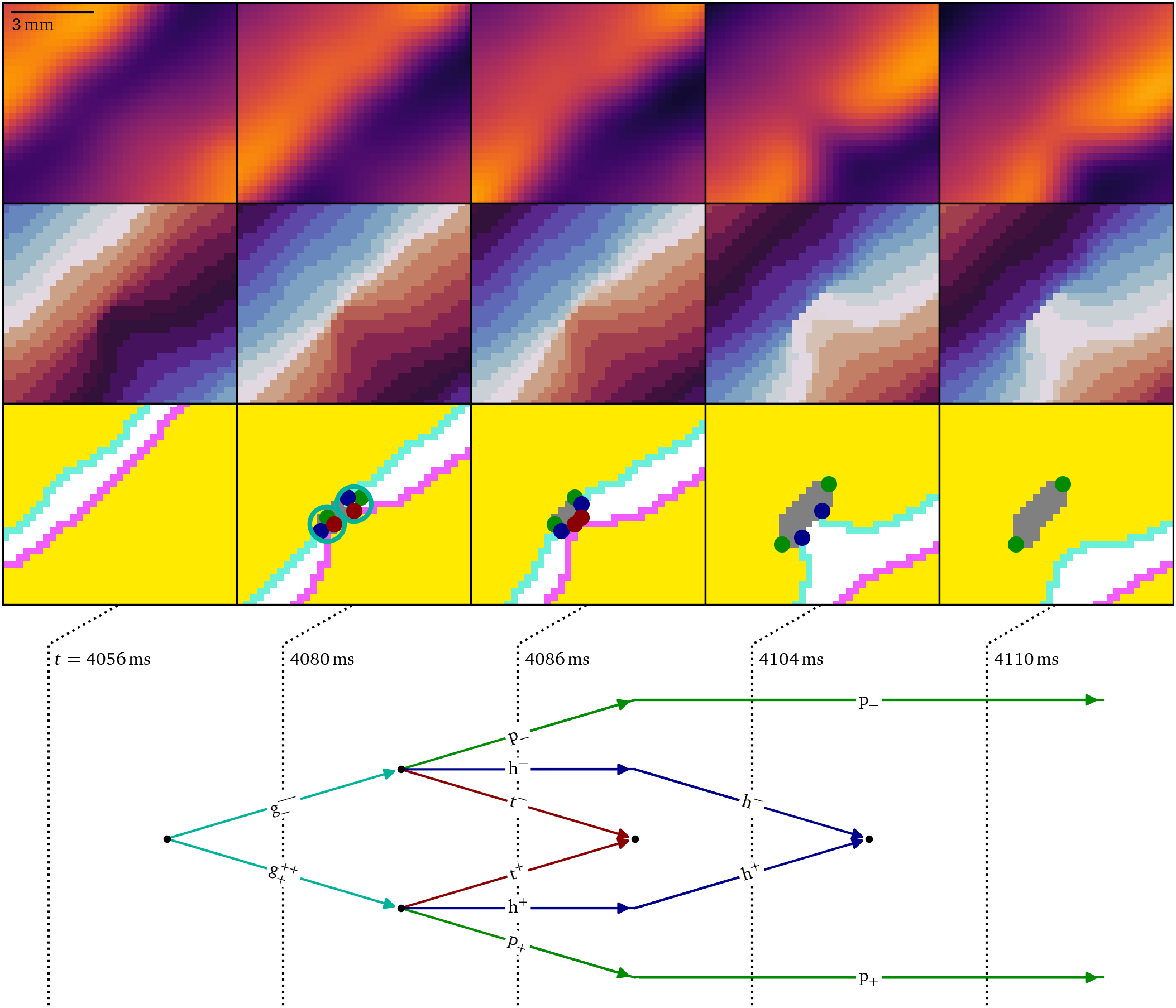}
    \caption[
        Creation of a persisting pair of pivot particles due to conduction block.
    ]{
        Creation of a persisting pair of pivot particles due to conduction block
        as the first step of figure-of-eight spiral formation in the optical mapping experiment,
        cf.~Fig.~\ref{fig:hiam}A.
        At first, when the wave front runs into the wave back,
        a pair of growth particles is formed, which then decay into a pivot, head, and tail each.
        The heads and tails annihilate, while the pivot particles persist.
    }
    \label{fig:hiam:init}
\end{figure*}

\begin{figure*}[tp]
    \centering
    \includegraphics[width=\textwidth]{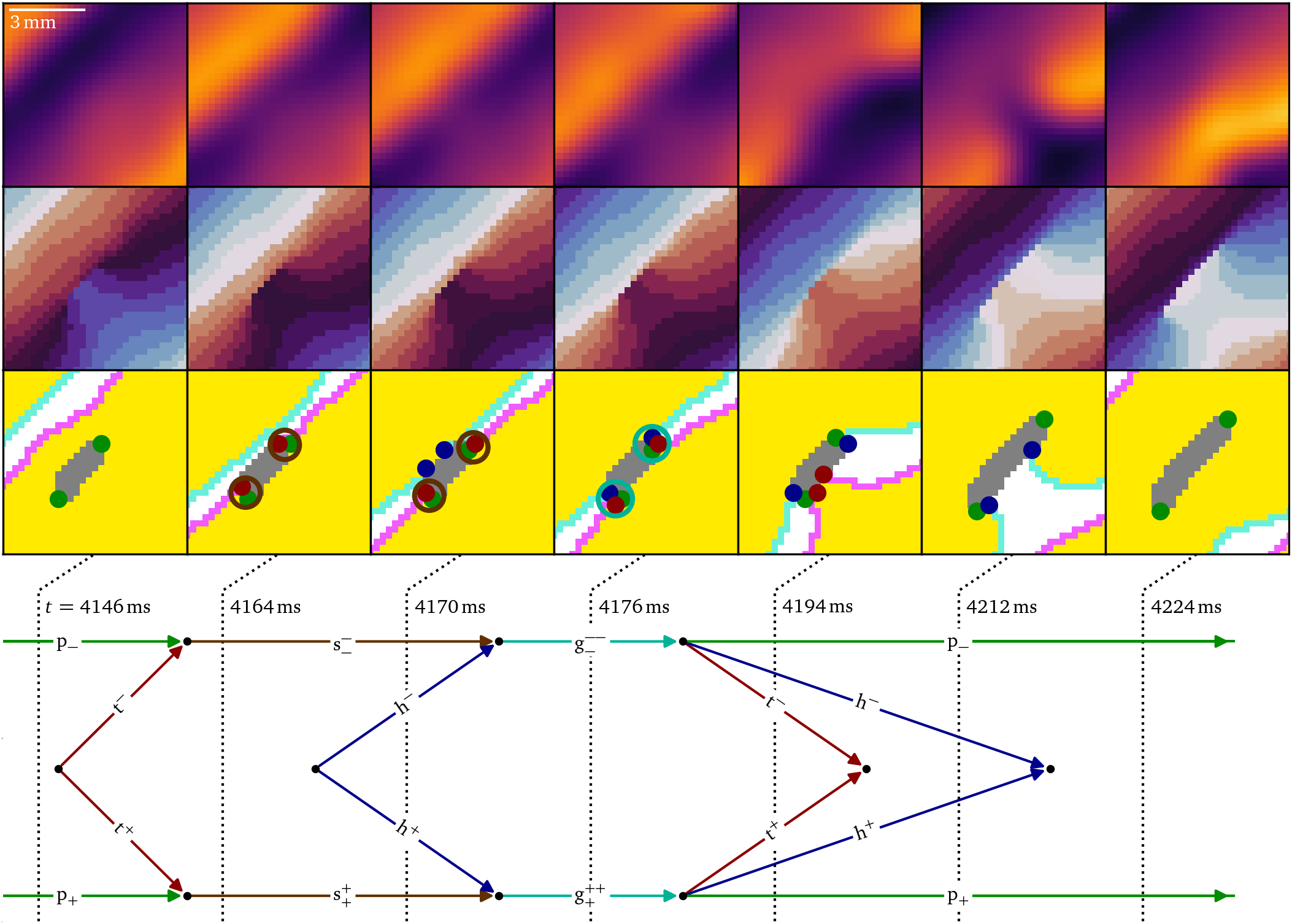}
    \caption[
        Accumulation of conduction block.
    ]{
        Accumulation of conduction block is the process of shrinking due to a wave back hitting a phase defect line and growing due to a wave front, which, in total, makes the conduction block grow.
        This process is observed another five times in the data during the following burst pacing pulses
        leading up to figure-of-eight spiral formation in the optical mapping experiment,
        cf.~Fig.~\ref{fig:hiam}B.
    }
    \label{fig:hiam:accumulation}
\end{figure*}

\begin{figure*}[tp]
    \centering
    \includegraphics[width=\textwidth]{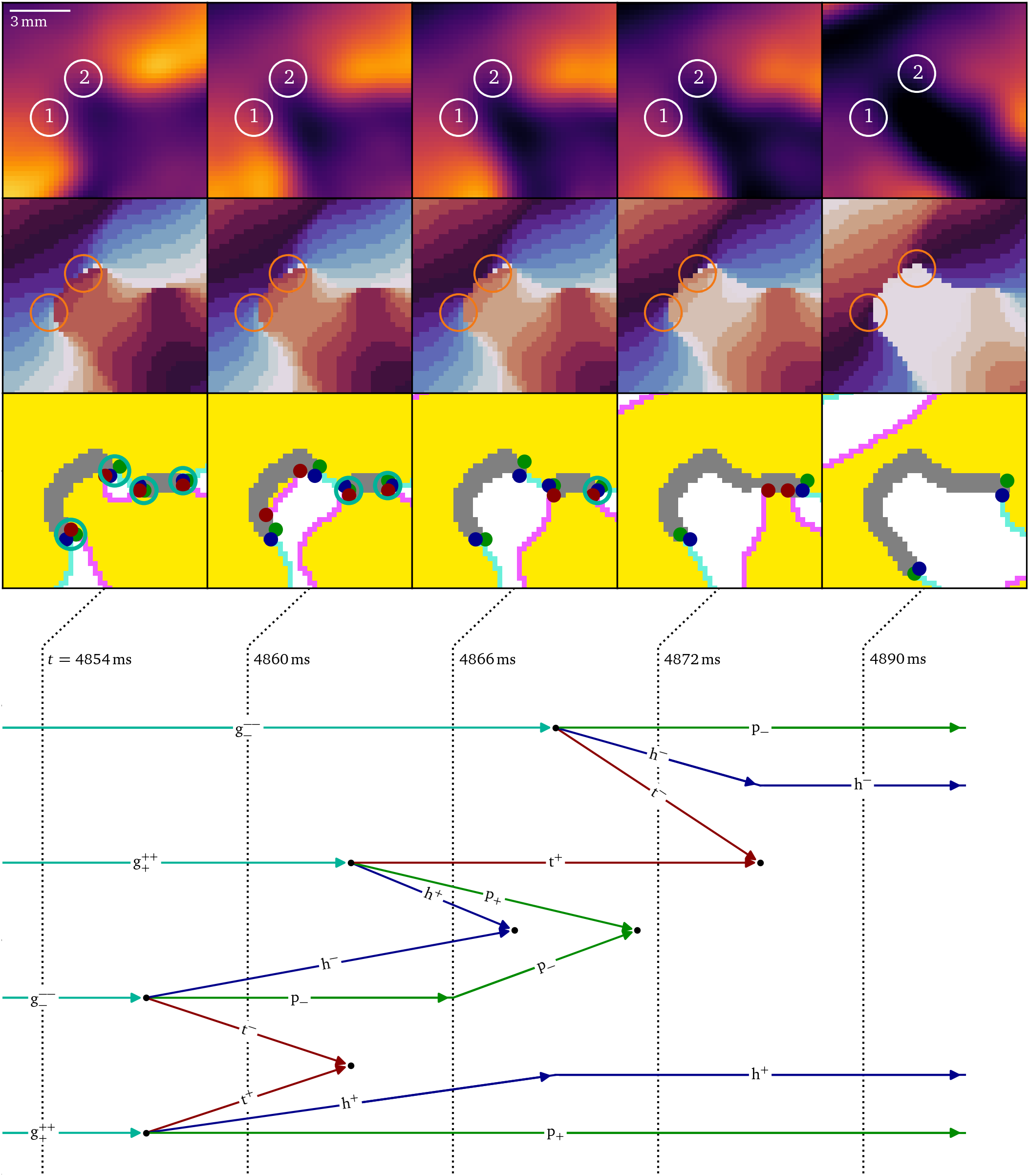}
    \caption[
        The merger of two conduction blocks.
    ]{
        The merger of two conduction blocks
        takes place when the two phase defects grow close enough such that
        their particles annihilate with each other,
        creating a larger phase defect line.
        This is another intermediate step of figure-of-eight spiral formation in the optical mapping experiment,
        cf.~Fig.~\ref{fig:hiam}D.
    }
    \label{fig:hiam:merge}
\end{figure*}

\begin{figure*}[tp]
    \centering
    \includegraphics[width=\textwidth]{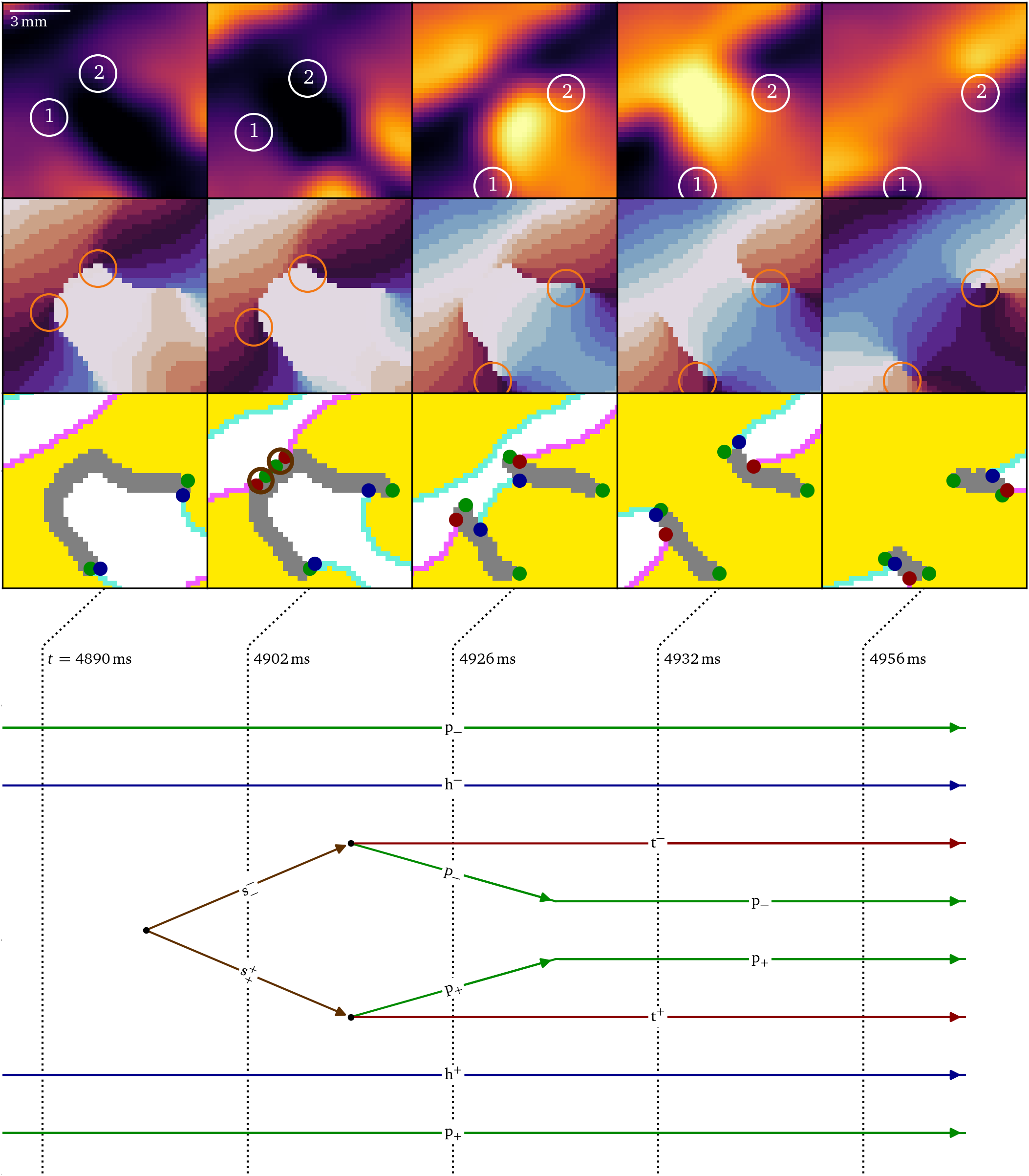}
    \caption[
        Splitting of a large conduction block.
    ]{
        Splitting of a large conduction block
        as the final step of figure-of-eight spiral formation in the optical mapping experiment,
        cf.~Fig.~\ref{fig:hiam}E.
        When a wave back meets the phase defect, a pair of shrink particles is formed,
        halving the U-shaped phase defect line with $Q=0$ and $P=0$.
        When the shrink particles decay, two rotor cores with $Q=\pm 1$ and $P=\pm 1$ are formed.
    }
    \label{fig:hiam:split}
\end{figure*}

The creation of a figure-of-eight spiral pair in a monolayer of human atrial myocytes \cite{Harlaar:2021}
is thoroughly analyzed via the quasiparticles and diagrams in  Figs.~\ref{fig:hiam}-\ref{fig:hiam:split}, see also the Methods section below.
By adding a voltage-sensitive dye to the cell culture, the local transmembrane voltage could be optically mapped \cite{Salama:1987}, which is shown here after normalization to $[0,1]$.
The culture was stimulated at a high frequency from the top left corner. After several transient wave breaks, i.e., conduction blocks, a pair of oppositely rotating spiral waves was formed.
The experiment presents an \emph{in vitro} realization of the onset of an arrhythmia, an event to be prevented in patients.

Fig.~\ref{fig:hiam} presents a diagrammatic overview of the figure-of-eight spiral pair formation,
while Figs.~\ref{fig:hiam:init},~\ref{fig:hiam:accumulation},~\ref{fig:hiam:merge},~\&~\ref{fig:hiam:split} present parts of this process in more detail in the same style as Fig.~\ref{fig:smooka_creation}.

As can be seen in Figs.~\ref{fig:hiam}A~\&~\ref{fig:hiam:init},
the initial conduction block produces a growth pair just before $t=\qty{4056}{\milli\second}$,
in the same way as in the \emph{in silico} case above,
which then break apart into a head, tail and pivot each.
The tails and heads annihilate and leave the persisting pair of pivots behind, marking the end points of the conduction block line.
Later on, a secondary conduction block forms in the same way, also producing a pivot-pair, cf.~Fig.~\ref{fig:hiam}C.

The process causing the the mere conduction block to develop into a dangerous figure-of-eight reentry,
which can be life-threatening in peoples' hearts,
can be seen in Figs.~\ref{fig:hiam}B~\&~\ref{fig:hiam:accumulation}:
After the wave back hits the phase defect region of the conduction block,
a tail pair is produced. Each tail then meets the persisting pivot particles to produce shrink particles that
begin to shorten the conduction block line at $t=\qty{4164}{\milli\second}$.
However, shortly afterwards at $t=\qty{4170}{\milli\second}$,
the wave front also hits the phase defect,
leading to a pair of heads, that combine with the shrinks to growth particles.
Driven by this pair, the conduction block line grows again,
until the growth particles decay into a head, tail and pivot each.
Just after $t=\qty{4194}{\milli\second}$, the tails annihilate,
as well as the heads just after $t=\qty{4212}{\milli\second}$,
while again two pivot particles persist.
As the growing outweighs the shrinking in this case, in total, the phase defect length accumulated.
The distance between the pivots increased from $\qty{2.850438562747845 +- 0.5}{\milli\meter}$ at $t=\qty{4146}{\milli\second}$ to $\qty{4.650268809434569 +- 0.5}{\milli\meter}$ at $t=\qty{4212}{\milli\second}$.
This process repeats six times, each time making the phase defect longer.

At $t=\qty{4854}{\milli\second}$, also the second phase defect has formed, and both will merge, as visualized in Figs.~\ref{fig:hiam}D~\&~\ref{fig:hiam:merge}. The merging process starts when
both defects grow with a pair of $\mathrm{g}^+$ and $\mathrm{g}^-$ each.
Then, the growth particles decay in triplets of head, tail and pivot and the tail-pairs of both phase defect sites
annihilate just as in the accumulation stage (Figs.~\ref{fig:hiam}B~\&~\ref{fig:hiam:accumulation}).
Meanwhile,
the $\mathrm{h}^-$ of the large phase defect and the $\mathrm{h}^+$ of the smaller phase defect annihilate with each other just after $t=\qty{4866}{\milli\second}$,
as well as their respective $\mathrm{p}^-$ and $\mathrm{p}^+$, merging the two phase defects into one larger U-shaped phase defect line.
At $t=\qty{4890}{\milli\second}$, a pair of pivots and heads each with opposing charges remain.
The pivots now have a distance of $\qty{6.7175144212722016 +- 0.5}{\milli\meter}$.

Finally, in Figs.~\ref{fig:hiam}E~\&~\ref{fig:hiam:split},
it can be seen how the recovery of the initial zone of contact creates a shrink pair,
which causes the splitting of the conduction block line just before $t=\qty{4902}{\milli\second}$.
Shortly after that, each shrink particle decays into a pivot and tail.
At $t=\qty{4932}{\milli\second}$, the system has evolved to a state with two nearly parallel conduction block lines, i.e., phase defects that comprise an effective functional isthmus around which a figure-of-eight rotor pair can revolve.
Each of the phase defects have a head, a tail, and two pivot particles, all with the same charge.
The phase defect with a clockwise rotating rotor has $Q=-1$ and $P=-1$, while the counter-clockwise rotor has $Q=+1$ and $P=+1$.
From $t=\qty{4956}{\milli\second}$, the pivots remain in place and the head and tail propagate along the phase defect line.
Zooming out, these particles behave like a pair of cores.

% time 4080 to 4902: spanning 820ms

In brief, the quasiparticle framework here reveals that the establishment of the figure-of-eight re-entry is a multi-step process that takes over $\qty{800}{\milli\second}$
to take place.
The process features tight interplay between dynamically changing conduction blocks and wave front ends (heads).
The formalism unifies the concepts of rotors and conduction block lines, and reveals how an initially flat conduction block line first takes a U-shape ($t=\qty{4890}{\milli\second}$)
and then becomes a functional isthmus.
The details of this interaction are not visible in the classical phase singularity analysis, which would only reveal the creation, vanishing, and recreation of a phase singularity pair.

\section*{Discussion}

\subsection*{An extended quasiparticle viewpoint on excitation}

The above analysis and examples show that complex excitation patterns feature certain preserved special points. Noteworthily, these are critical points in space, rather than critical points of evolution equations, which also satisfy topological rules\cite{glass_topological_1975}. While tips, pivots and cores have been observed before, the growth and shrink quasiparticles emerge from the framework itself. This situation is to a certain degree reminiscent of the `Standard Model' in physics, which is based on a small set of fundamental particles \cite{mann:2010} and explains almost all experimental results in particle physics. The theoretical framework presented in this work gains its power from simple geometrical arguments, as the heads and tails are points where three zones meet: excited, unexcited, and phase defect. The concept of `head' itself has been implicitly used before, e.g. in Krinsky's quantitative theory of reverbators \cite{krinsky_fibrillation_1968,Krinsky:1992} for inhomogeneous domains. Our framework is much wider valid, as it can describe long and short-lived rotors, wavelets and conduction blocks in homogeneous and inhomogeneous excitable media.

\subsection*{Cardions}

In physics, it is customary to give localised and preserved structures a name with the suffix `-on', with the first part of the name referring to the nature of the system of the (quasi-)particle\cite{mw:on}. Examples include not only most particles in the Standard Model (proton, electron, baryon, fermion, photon) but also other moving structures such as phonons, solitons and excitons. In analogy, we here propose to name the here-identified quasiparticles as `cardions', as they were first observed in cardiac excitation patterns.

\subsection*{Dynamical transitions}

The motivation for this study was to investigate the interactions between short-lived rotors and approaching wave fronts during arrhythmogenesis \cite{Clayton:2005}. We go beyond the observation that apparent phase singularities co-locate with conduction blocks \cite{Rodrigo:2017,Arno:2021}, by looking at a finer spatial scale. We discovered multiple interactions of the cardions in the data and represented them schematically. Our initial analyses show new mechanistic insight, like the dynamical isthmus formation in Fig. \ref{fig:hiam:split}. It can be expected that the automated construction of those diagrams for larger fibrillation domains (see Fig. \ref{fig:hiam}) will allow an enhanced classification and characterisation of those patterns.

\subsection*{Links between particle physics and cardiology}

The quasiparticle viewpoint allows us to introduce more concepts from particle physics in the world of excitable media. First, note that the terms `creation' and `annihilation' also applied to the classical viewpoint \cite{Gray:1998,Clayton:2005}. There, a single quasiparticle was present, i.e., the spiral wave centre (phase singularity), bearing a positive or negative charge ($Q=\pm 1$) depending on its rotation sense. In our viewpoint, creation and annihilation also occurs with heads, tails and pivots.  Since cardiac tissue is, in reality, non-homogeneous, a wave front will often create a head pair near obstacles, e.g. areas of fibrosis, which annihilates shortly afterwards. Second, the difference between linear-core and circular-core rotors follows from the mutual interaction between pivots, which can be expressed with spiral wave response functions \cite{Biktasheva:2003,Marcotte:2016,Dierckx:2017}. If nearby pivot charges of the same charge attract each other, they form a core particle. Conversely, when they repel each other, a linear-core rotor emerges. However, such a phase defect cannot grow larger than the width of a travelling pulse in the medium, since otherwise, the defect line will break by the creation of a new $\mathrm{p}_+$, $\mathrm{p}_-$ pair. Therefore, pivot particles are in our opinion subjected to confinement, analogous to quarks in quantum chromodynamics \cite{wilson_confinement_1974}. This confinement may partially explain why the here-identified quasiparticles remained unnoticed during simple dynamical regimes, i.e., at time intervals where no new vortices or conduction blocks form or disappear.

Finally, we note that in Fig. \ref{fig:intro}B, there are two apparent symmetries.  First, if one exchanges excited with unexcited regions,  wave fronts become wave backs, heads become tails and vice versa.  This corresponds to flipping the voltage axis around the value $V_*$ from \eqref{phi_act}. Second, one can reverse the arrow of time, which preserves $\mathrm{U}$ and $\mathrm{E}$ regions, but interchanges wave backs with fronts and heads with tails.  However, when looking at the dynamics, there is a clear distinction between heads and tails. As heads lie on a wave front, they move at the conduction velocity in the medium. Since the repolarisation of the tissue occurs at a fixed delay after the depolarisation, the wave back is a phase wave, and wave backs and tails can move at any speed. Also, the medium can remain in the unexcited state for a unlimited time but only a limited time in the excited state. This effective breaking of symmetry also manifests in the growth and decay states. In the cases we have yet observed in simulations and experiment, the growth of a phase defect line each time involved the three different quasiparticles. Namely, a conduction block arises when a wave front (h) hits a wave back (t) and in the process, a phase defect is formed (p). Conversely, the vanishing of a phase defect only requires that the phases at both sides return to full recovery. We took the threshold of the wave back to mark the state where a standard impeding wave front can re-excite the tissue. Then, only a tail and pivot particle are needed. In case of a heterogeneous medium, we anticipate that also other situations can occur, but these fall outside the scope of our present investigation. Noteworthily, a statistical analysis of occurring processes in the topological approach of Marcotte~and~Grigoriev~\cite{Marcotte:2017} also demonstrated asymmetry between favoured processes in the creation and annihilation processes of spiral cores.

%Intriguingly, the growth of a conduction block line requires 3 quasiparticles, while the

\subsection*{Limitations}

The parallels drawn here between biological excitation and theoretical physics are only at the conceptual level. Feynman diagrams in physics are representations of a path integral, allowing to structure tedious calculations. To use the diagrams in excitable context for quantitative calculations and predictions could be a next research step. Even then, the mathematical implications will be different from the physics case. For this reason, we refer to the diagrams here consistently as `Feynman-like'. Likewise, the terms of confinement, quasiparticle interaction, standard model and symmetry breaking are powerful analogues when used in complex systems, but will at a more detailed level have different meaning than their original physics use.

The local state of a cell was here assumed to lie along a cycle, such that it can be labelled with a single phase variable $\varphi(\vec{r},t)$. Capturing specific memory effects, such as alternans \cite{Karma:1994} or after-depolarisations in cardiac tissue \cite{weiss_early_2010} will require an extension of the formalism, e.g. using phase-amplitude equations\cite{wilson_phase-amplitude_2020}.

We are also not certain if all possible cardions states have been described here. We already noted that phase defects can have side branches, and the `joints' where this happens will also be cardions, to be described elsewhere. In
Fig.~\ref{fig:hiam:merge}, co-localisation of a $\mathrm{h}$ and a $\mathrm{p}$ particle is seen in the frames. However, in our experience this state is transient and therefore not considered a bound state here. Nonetheless, it is possible that other regimes exist where $\mathrm{h}$-$\mathrm{p}$ pairs may travel together as a bound state.

The concept of $P$-charge was introduced in this work, and our understanding of it is currently incomplete. The $P$-charge is currently undefined before the first wave crosses the medium. The precise conditions under which it is conserved or not need to be further elucidated. For isolated, thin, conduction block lines, $P$-charge seems to be conserved: Basic features in the patterns observed are linear rotor cores ($P=\pm 1$) or mere conduction blocks ($P=0$).

%A related quantity that is perhaps useful can be obtained by looking at the direction tangential to the contour circling the phase defect. Say the contour $\mathcal{C}$ is followed such that the phase defect is to the left, and that it encloses an angle $\chi$ with the positive $X$-axis. Then let us introduce the unsigned $P$-charge:
% \begin{align}
%     \tilde{P} = \frac{1}{2\pi} \int d \chi = \frac{1}{2\pi} \int k ds.
% \end{align}
% here, $\frac{d\chi}{ds} = k$ is by definition the local curvature of the contour. For any closed curve, $\tilde{P}$ evaluates to $1$. Near a pivot point, the curve turns over $180^\circ$, such that there, $\tilde{P}=$

Finally, we have applied the framework here already on \emph{in silico} and \emph{in vitro} datasets. An obvious next step would be to also analyse surface patterns in cardiac tissues, and to post-process phase maps or LAT maps obtained in patients. We encourage the scientific community to work together on these steps and are providing basic numerical methods for this on GitLab, see below. We are currently working on automated algorithms to detect cardions in datasets.

\subsection*{Applications to cardiac arrhythmia}

The above examples revealed dynamics at a finer level than in the phase singularity picture. Conducting similar analyses on existing datasets during arrhythmias is likely to give new clues on underlying mechanisms. More specifically, counting the elementary interactions \cite{Marcotte:2017}, computing statistics of potentially large  diagrams, and deriving interaction laws between the quasiparticles could prove useful. The value of the $P$-charge for obstacles in the medium could also serve as a `distance' to the rotor regime, and thus quantify how far a regime is from arrhythmia, complementing other risk scores in literature \cite{arevalo_arrhythmia_2016}.

% to do:
% limitations of the correspondence!!
% application to other fields

The pivot sites, here identified as quasiparticles, are actively being investigated by clinicians as possible ablation targets \cite{seitz_af_2017}. We believe that the analysis of the clinical data and the pivot ablation within our framework could help to answer if, why, and when these points are appropriate ablation targets.

\subsection*{Outlook}

In this manuscript, we have exposed several new concepts, such as the pivot charge, different quasiparticles and a diagrammatic approach to better understand their interactions. Continuing research by ourselves and colleagues will be needed to elaborate these concepts.

Within the context of  continuous excitable media, several extensions are possible. Since we do not suppose an underlying evolution model, the analysis could also be performed to the surface of three-dimensional media. A next step is to analyse optical voltage mapping surface recordings of the heart muscle. Such analysis can be applied to any excitable surface pattern that is sufficiently sampled in space and time. Our original motivation is to perform predictive calculations in those patterns: Can one find the critical size of a wave break, or the timescale at which a rotor produces multiple wavelets? In three spatial dimensions, our preliminary findings\cite{Arno:2024b} indicate that the cardions become string-like objects\cite{Verschelde:2007}. As such, cardions in three dimensions will refine the concept of a rotor filament \cite{Winfree:1973,Clayton:2005}, and may be used in the future to further investigate the three-dimensional organisation of turbulence in complex systems.

\section*{Conclusion}

We presented a conceptual framework revising the classical theory of excitable media, intentionally designed for complex regimes: short-lived pivoting motion and prominent conduction blocks. By classifying the endpoints of wave fronts, wave backs, and conduction blocks as charged quasiparticles, Feynman-like diagrams can be created from phase maps or local activation times. We believe this framework has the potential to become a useful analysis tool in excitable media, with applications within and beyond
the cardiac electrophysiology context.

\section*{Methods}\label{app:methods}
\subsection*{\emph{In silico} data generation}

The synthetic data from Fig.~\ref{fig:intro}A were obtained by Euler-forward stepping of the Bueno-Orovio-Cherry-Fenton model \cite{BuenoOrovio:2008}, with finite-differences grid size $\qty{0.5}{\milli\meter}$ and time step $\qty{0.1}{\milli\second}$ in a biventricular heart geometry using an S1S2 stimulation protocol in the Ithildin~framework~\cite{kabus2024ithildin}.
The data from Fig.~\ref{fig:intro}B are obtained in the same way but for a square piece of tissue with grid size $\qty{0.3}{\milli\meter}$.
For the fibrillation-like data from Fig.~\ref{fig:smooka_creation},
the Karma~model~\cite{Karma:1993, Karma:1994} modified according to Marcotte~\emph{et~al.}\cite{Marcotte:2017} was used,
with step size $\qty{1}{\milli\meter}$ and time step $\qty{0.1}{\milli\second}$.

\subsection*{\emph{In vitro} data generation}

Monolayers of fully functional human atrial myocytes were generated as described in detail by Harlaar~\emph{et~al.}\cite{Harlaar:2021}. A voltage-sensitive dye was added to the culture, after which a real-time recording can be made of the intensity of emitted light, which is a measure of the local transmembrane potential \cite{Salama:1987}. The recording has a resolution of $100 \times 100$ pixels with pixel size $\qty{0.25}{\milli\meter}$. The sampling time between frames was $\qty{6}{\milli\second}$.

\subsection*{Data analysis}

Local activation times were converted to phase following \eqref{phi_lin} with time constant $\tau_0 = \qty{110}{\milli\second}$ for the \emph{in vitro} data and $\qty{70}{\milli\second}$ for the \emph{in silico} data. We use the phase values $\varphi_\mathrm{F} = \qty{0}{\radian}$ and $\varphi_\mathrm{B} = \qty{5}{\radian}$. Phase defects are identified following \eqref{Z} with the threshold $\rho_* = 0.2$. The positions of tips, heads, tails, pivots and compound particles were drawn manually on the resulting figures.

\subsection*{Data availability}

The simulation output and pre-processed optical voltage mapping data
we have used for this article can be found on \href{https://doi.org/10.5281/zenodo.13379783}{Zenodo (DOI: 10.5281/zenodo.13379783)}.
This repository also contains the scripts and Python modules to generate the
figures found in this article.
The source code for
the Ithildin finite-differences solver\cite{kabus2024ithildin} (\href{https://gitlab.com/heartkor/ithildin}{https://gitlab.com/heartkor/ithildin}),
its Python module\cite{kabus2022numerical} (\href{https://gitlab.com/heartkor/py_ithildin}{https://gitlab.com/heartkor/py\_ithildin}),
and a Python module to create the Feynman-like diagrams (\href{https://gitlab.com/heartkor/pdl-feynman}{https://gitlab.com/heartkor/pdl-feynman})
can also be found on GitLab.

% \bibliography{references}

\section*{Acknowledgements}

The authors are grateful to D.A. Pijnappels, A.A.F. de Vries and N. Harlaar for collecting and sharing the hiAM dataset. The authors thank J. Ector and T. De Coster for helpful discussions. The authors thank A.A.F. de Vries for useful comments on the manuscript. The authors thank A. Gobeyn for coining the term `cardion' for the collection of quasiparticles.

\section*{Author contributions statement}

L.A.: conceptualisation, methodology, software, validation, formal analysis, investigation, data curation, writing -- original draft, writing -- review \& editing, visualisation.
D.K.: conceptualisation, methodology, software, validation, formal analysis, investigation, writing -- original draft, writing -- review \& editing, visualisation.
H.D.: conceptualisation, methodology, resources, writing -- original draft, writing -- review \& editing, supervision, project administration, funding acquisition.
Correspondence and requests for materials should be addressed to Hans Dierckx.

\section*{Competing interests}

The authors declare no competing interests.

\section*{Additional information}

L. Arno is funded by a FWO-Flanders
fellowship, grant 117702N. D. Kabus is supported by KU Leuven grant GPUL/20/012. H. Dierckx was supported by KU Leuven grant STG/019/007.

\end{document}